\documentclass[prb,twocolumn,aps,showpacs]{revtex4}
\usepackage{graphicx,color}
\usepackage{latexsym,amssymb}

\def\be{\begin{equation}}
\def\ee{\end{equation}}

\def\bc{\begin{center}}
\def\ec{\end{center}}
\def\ds{\displaystyle}

\def\ko{k^{\rm O}}
\def\KO{K^{\rm O}}

\def\ku{k^{\rm U}}
\def\KU{K^{\rm U}}

\def\ks{k^{\rm S}}
\def\KS{K^{\rm S}}

\def\gs{\gamma^{\rm S}}

\def\ml{\ell}

\begin{document}

\title{Two-dimensional  electron systems  beyond the diffusive  regime}

\author{P. Marko\v{s}}
\affiliation{%
Department  of Physics FEI, Slovak University of Technology,   812\,19 Bratislava, Slovakia}

\begin{abstract}

Transport properties of disordered electron system can be characterized by 
the conductance, Lyapunov exponent, or level spacing. Two
additional parameters, $K_{11}$ and $\gamma $ were introduced recently which measure the non-homogeneity
of the spatial distribution of the electron inside the sample. 
For the orthogonal, unitary  and symplectic  two dimensional disordered models, 
we investigate numerically the system size dependence of these parameters
in the diffusive and  localized regime.
Obtained size and disorder dependence of $K_{11}$ and $\gamma$ is in agreement with  with single parameter transport theory.
In the localized regime, $\gamma\to 0$ independently on the physical symmetry of the model.
In the diffusive regime, $\gamma$ equals to the symmetry parameter $\beta$.
For the symplectic model we analyze the size dependence of $\gamma$ 
in the critical region of the metal-insulator transition and found 
 the non-universal critical value  $\gamma_c$. 
\end{abstract}

\pacs{73.23.-b, 71.30.+h, 72.10.-d}

\maketitle

\section{Introduction}

Transport of electrons  through  disordered  structures 
offers a broad variety of interesting  universal phenomena.
\cite{lee,McKK-93} 
With increase of the strength of the disorder the character of the transport 
changes from the ballistic to diffusive up to the insulating, where all electrons are
localized. \cite{Anderson-58}.

In the limit of weak disorder
(diffusive regime) the transport  can be studied  analytically  using, for instance,
 the Dorokhov Mello Pereyra Kumar (DMPK) equation 
\cite{dmpk} 
the Green's function analysis\cite{ucf} or random matrix theory.
\cite{pichard-nato,been}
The existence of the metal-insulator transition in two  and three 
dimensional  models \cite{AALR,evers}  is a strong motivation to
construct an analytical theory of the transport beyond the diffusive regime
\cite{somoza-prl,garcia}.
Also, numerical data for the localized regime 
\cite{markos2,acta,prior,somoza,Z}
show that, contrary to theoretical expectation, 
the distribution of the logarithm of the conductance is never Gaussian for disordered
systems in higher dimension.  
Therefore, a general transport   theory   must explain how the dimension of the system and
physical symmetry of the model \cite{evers} influence the ability of electron to move 
through the sample.

The most elaborated analytical description of the transport in strongly disordered structures
is  based on the generalized DMPK equation (GDMPKE).  \cite{mk}
The theory takes into account that the spatial distribution
of electrons in the regime of localization  is not homogeneous.  The last was confirmed
by numerical simulations in  Ref. \cite{etopim,prior} 
In GDMPK,  the non-homogeneity of electron distribution is measured by a large number of 
 parameters $K_{ab}$ (defined later); however, only two of them,
$K_{11}$ and $\gamma=2K_{12}/K_{11}$ are decisive for the transport.\cite{mmwk}

The GDMPKE is not exactly solvable, but approximate analytical solution 
for 3D disordered systems \cite{mmwk,MMW,douglas} agrees very well with numerical data. 
Numerical solution of GDMPKE \cite{Brnd-2007} confirmed that it correctly describes disordered 
orthogonal systems and that parameters $K_{ab}$  depend on the dimension of the system.

Detailed numerical analysis of parameters $K_{11}$  and $\gamma$ in three dimensional model 
 was performed in \cite{MMW}. 
The aim of this paper  is to investigate how these parameters depend on the physical symmetry in
two dimensional (2D) models. 
We present numerical data for the parameters $K_{11}$  and $\gamma$ for  
the  orthogonal model  (O), unitary (U)  and two  symplectic (S)\cite{EZ,Ando-89}
models
in diffusive and insulating regime. For the S models, we also study the  behavior of both
parameters in the critical regime of the metal-insulator transition.

\section{Generalized DMPK equation}

Consider  a disordered system of the length $L_z$ connected to
two semi-infinite ideal leads with $N$ open channels.  
Transmission parameters are given by the transfer matrix, 
which can be written in general form as \cite{dmpk}
\begin{equation}\label{one}
T=\left(\matrix{ u & 0  \cr 0 & u' \cr }\right) \left(\matrix{
\sqrt{1+\lambda} & \sqrt{\lambda}   \cr \sqrt{\lambda}   &
\sqrt{1+\lambda} \cr }\right)\left(\matrix{ v & 0  \cr 0 & v'
\cr }\right).
\end{equation}
In Eq. (\ref{one}), 
$u,v$ are $N \times N$ matrices, and 
$\lambda$ is a diagonal matrix, with positive elements $\lambda_a, a=1,2, ...N$.
In systems with time reversal symmetry, matrices $u'$ and $v'$ can be represented in terms of
$u$ and $v$.  \cite{MelloPichard}
For the orthogonal system, $u' = u^*$  and $v'=v^*$.  
For the symplectic symmetry, the scattering depends on the spin of the electron; the elements 
of matrices $u$ and $v$ are $2\times 2$ matrices which fulfill the symmetry relations
\cite{pichard-nato,MelloPichard}
\be
u' = k u^* k^T, ~~~~~
v' = k v^* k^T, ~~~~~
k= \left(
\begin{array}{rr}
0 & -1\\
1 &  0
\end{array}
\right).
\ee
Statistical variables $u$, $v$ and $\lambda$  contain entire information about the transport.
In the weak disorder limit,\cite{dmpk} the conductance $g$ (in units of $2e^2/h$)
is completely determined by eigenvalues $\lambda_a$.
\cite{SE,pichard-nato}
\be\label{gg}
g =   \sum_{a=1}^N\ds{\frac{1}{1+\lambda_a}}.
 = \sum_{a=1}^N \displaystyle{\frac{1}{\cosh^2 x_a/2}}
\ee
In the last equation, we used the parametrization $\lambda_a=(\cosh x_a-1)/2$. 

The probability distribution of $\lambda$s can be found as a solution of the DMPK equation.\cite{dmpk}
The generalization of the DMPK for the orthogonal symmetry class, was done by
 Muttalib and Klauder \cite{mk} who introduced new parameters,
$K_{ab}$  which  characterize the spatial distribution of the electron in 
the disordered sample. The generalized DMPK equation reads
\cite{mk}
\be\label{gdmpk}
\frac{\partial {p}_{L_z}(\lambda)}{\partial (L_z/\ml)}
=\frac{1}{{J}}\sum_a^N
\frac{\partial}{\partial\lambda_a}\left[\lambda_a(1+\lambda_a)K_{aa}
{J}\frac{\partial {p}}{\partial \lambda_a}\right],
\ee
where $\ell$ is the mean free path, and
\be
{J}\equiv\prod_{a<b}^N|\lambda_a-\lambda_b|^{\gamma_{ab}},
~~~~~~~\gamma_{ab}\equiv\frac{2K_{ab}}{K_{aa}}.
\ee
This equation can be simplified when all $K_{aa}$ are approximated by $K_{11}$
and $\gamma_{ab}\approx \gamma$ for all $a$, $b$ ($a\ne b$).  This approximation was confirmed by
numerical  work \cite{MMW,Brnd-2007}.

Although the conductance is still given by Eq. (\ref{gg}), it becomes implicitly a function of the
spatial distribution of the electron.

\section{Models}

In  numerical work, disordered sample is represented by two dimensional (2D) square disordered lattice
of the size $L\times L$.
The  orthogonal 2D model with on-site  disorder is defined by  the Hamiltonian 
\be\label{ham}
\begin{array}{lcl}
{\cal H} =&~~& W \sum_{xz}\epsilon_{xz}c^\dag_{xz}c_{xz} \\ 
&+& V_\perp\sum_{xz}c^\dag_{x+a,z}c_{xz}+c^\dag_{xz}c_{x+a,z}\\
&+& V_\parallel\sum_{xz}c^\dag_{x,z+a}c_{xz}+c^\dag_{xz}c_{x,z+a}
\end{array}
\ee
Here, $a$ is the lattice spacing, 
$\epsilon_{xz}$ are random energies from the box distribution, $|\epsilon_{xz}|<1/2$,
$W$ measures the strength of the disorder and $V_\parallel\equiv 1$ defines the energy scale.
To avoid closed channels in leads, we use  $V_\perp/V_\parallel =t<1$.
\cite{comment}  In what follows
we consider $t=0.9$,  the energy of the electron $E=0.01$. 
With $a\equiv 1$, we identify the number of channels
\be
N\equiv L.
\ee

It is generally accepted \cite{AALR,McKK-93} that only localized regime exists in the model
when the size of the system $L\to\infty$
(the critical disorder $W_c=0$). Nevertheless, diffusive transport is observable 
for sufficiently weak disorder and small sample size.\cite{acta}

The second model of interest is  the symplectic model   with spin dependent hopping. Here, 
the hopping of electron from one site to the neighboring one can be 
accompanied by the change of the sign of the spin and
$V_\parallel$,
$V_\perp$  become   $2\times 2$ matrices. In numerical simulations, we study  the 
Ando model with hopping  hopping terms
\begin{equation}
V_\perp=
t\left(\matrix{
V_1\hfill &-V_2\hfill\cr
V_2\hfill & V_1\hfill\cr
}\right),~~~~
V_\parallel=
\left(\matrix{
V_1\hfill &-iV_2\hfill\cr
-iV_2\hfill & V_1\hfill\cr
}\right).
\label{ando}
\end{equation}
The spin-orbit coupling is characterized by the 
parameter $S=V_1$ and $V_1^2+V_2^2=1$. In this paper, $S=0.5$. 
We also study the Evangelou-Ziman (EZ) model \cite{EZ} 
which uses the  random hopping matrices  $V$:
with  help of three independent random variables, $t^x,t^y,t^z$, 
distributed uniformly in interval $(-\mu/2,\mu/2)$
\begin{equation} \label{Ziman}
V_\perp = V_{xz,x+az}=
t \left(\matrix{
1+it^z\hfill &-t^y+it^x\hfill\cr
t^y-it^x\hfill & 1-it^z\hfill\cr
}\right),
\ee
and
\be
V_\parallel = V_{xzxz+a}=
\left(\matrix{
1+it^z\hfill &-t^y+it^x\hfill\cr
t^y-it^x\hfill & 1-it^z\hfill\cr
}\right),
\end{equation}
and  consider  $\mu=1$.

Both Ando and EZ model exhibit  the metal-insulator transition when the disorder $W$ reaches the
 critical value $W_c$. \cite{EZ,Ando-89}
Owing to the  anisotropy of our models, the critical disorder differs  from that obtained
in previous works.\cite{EZ,jpa} 
We found $W_c\approx 5.525$ for the Ando model and $W_c=6.375$ for the EZ model.

The 2D model with external magnetic field $B$ can be obtained by including 
the Peierls hopping term  $V_\perp = t\exp ix\alpha$,
$\alpha =  (e/\hbar) Ba^2$ into the Hamiltonian  (\ref{ham}).

\section{The matrix $K$}\label{kaka}

The matrix $K_{ab}$ is defined in terms  of higher moments of the matrices $v$:
\be\label{two}
K_{ab}\equiv \langle k_{ab}\rangle 
\ee
Here, $\langle\dots\rangle$  represents an  ensemble average.

For the orthogonal system, the matrix $\ko_{ab}$ is defined as \cite{mk} 
\be
\ko_{ab} = \sum_{\alpha=1}^L|v_{\alpha a}|^2|v_{\alpha b}|^2.
\ee
In the diffusive regime,\cite{dmpk}
\be\label{dmpk-q1do}
\KO_{ab} = \ds{\frac{1+\delta_{ab}}{L+1}}.
\ee
For the unitary models, 
\be
\ku_{ab} = \sum_{\alpha=1}^L|v_{\alpha a}|^2|v'_{\alpha b}|^2
\ee
and
\be\label{dmpk-q1du}
\KU_{ab} = \ds{\frac{1}{L}}.
\ee
For the systems with symplectic symmetry
the matrix $\ks_{ab}$ is given \cite{MC,Mello}
\be\label{ks}
\ks_{ab} =  \sum_{\alpha=1}^L v^\dag_{\alpha a} v^*_{\alpha b}\overline{v}_{\alpha b}v_{\alpha_a}.
\ee
In this equation, the $2\times 2$ 
matrices $v^\dag$, $v^*$ and $\overline{v}$ are defined in terms of the matrix $v$:
\be
v=
\left(
\begin{array}{rr}
              v_{11} & v_{12} \\
              v_{21} & v_{22}
\end{array}
\right)
~~~~~~~v^\dag=
\left(
\begin{array}{rr}
              v_{11}^* & v_{21}^* \\
              v_{12}^* & v_{22}^*
\end{array}
\right),
\ee
and
\be
v^*=
\left(
\begin{array}{rr}
              v_{11}^* & v_{12}^* \\
              v_{21}^* & v_{22}^*
\end{array}
\right),
~~~~~~\overline{v}=
\left(
\begin{array}{rr}
               v_{22} & -v_{12} \\
              -v_{21} & v_{11}
\end{array}
\right).
\ee
In the diffusive regime, $\KS_{ab}$ is degenerated diagonal matrix\cite{MC} with
diagonal elements
\be\label{dmpk-q1ds}
\KS_{ab} = \ds{\frac{2 - \delta_{ab}}{2L-1}}.
\ee
Our numerical results discussed in Sect. \ref{results} confirm that the same holds for any disorder
strength. 

From Eqs. (\ref{dmpk-q1do},\ref{dmpk-q1du},\ref{dmpk-q1ds}) it follows that
\be
\sum_b \KO_{ab} = \sum_b\KU_{ab} = \sum_b \KS_{ab} = 1.
\ee
We use these relations to test the  numerical accuracy of our results.

\section{Results}\label{results}

We consider square samples of the size $L\times L$ attached to two semi-infinite 
ideal leads. The size $L$ increases from $L=10$
to $L=256$ (S model) up to $L=600$ (O model). 
For each value of $L$ and $W$,
we analyze the statistical ensemble of typically $N_{\rm stat}\sim 10^4$ 
samples ($N_{\rm stat}\sim 1000 - 4000$ for the  largest system size).
In numerical calculation, the sample and leads are represented by  the $2N\times 2N$ 
transfer matrix $M$ and $M_0$, respectively.\cite{xx}
Following \cite{PMcKR}, the conductance is given as a trace of matrices 
$L^+MR^+$ and $L^-MR^-$, where $R^{+-}$ ($L^{+-}$) are $N\times 2N$ ($2N\times N$) matrices 
composed  of right and left eigenvalues of $M_0$, respectively. The upper index $^+$ ($^-$) 
indicate the direction of the propagation through the sample. Comparing with Eq. (\ref{one}) we find
\be
L^+MR^+ = v(1+\lambda)^{-1}v^\dag
\ee
and
\be
L^-MR^- = {v'}^{\dag}(1+\lambda)^{-1}v'.
\ee
Thus, eigenvalues $\lambda_a$ can be obtained numerically by diagonalizing
of the  matrices $L^+MR^+$ and $L^-MR^-$. Matrices $v$ and $v'$ consist 
of corresponding eigenvectors.
Details of numerical method are given in Ref. \cite{MMW}
Mean values, $K_{11}$ and $K_{12}$ were  calculated as an average over
the statistical ensemble
\be K_{ab} = \ds{\frac{1}{N_{\rm stat}}}\sum_{i=1}^{N_{\rm stat}} k_{ab}^{(i)}.\ee
Obtained data for $k_{ab}$ were also used for the calculation of probability
distributions.

As noted in Section \ref{kaka}, $\KS_{ab}$ are   $2\times 2$ matrices.
Numerical data confirm that, with the relative accuracy of $10^{-3}$, these matrices remain
diagonal degenerate
for each value of the disorder and all size of the system.

\begin{figure}[t!]
\includegraphics[width=6.0cm]{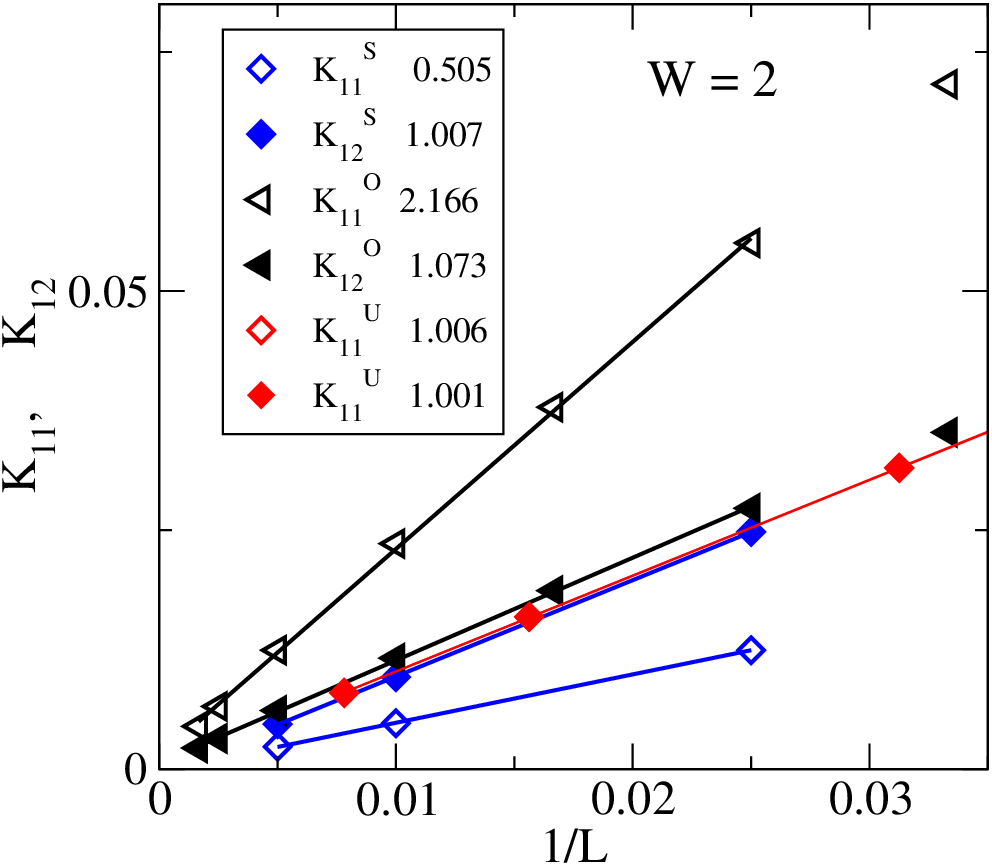}
\caption{
(Color online) $K_{11}$  and  $K_{12}$  
as a function of the system size for square lattice $L\times L$
for the symplectic (S) and orthogonal (O) s and unitary (U) systems. The disorder $W=2$.
Solid lines are linear fits with slopes given in the legend.
}
\label{K11-all}
\end{figure}

\subsection{Diffusive regime}

We first verify the prediction of the DMPK equation for the diffusive regime.
In Fig. \ref{K11-all}  we show the $L$ dependence of parameters $K_{11}$  and $K_{12}$
for the orthogonal and symplectic system with disorder $W=2$. The system is in the diffusive regime 
(the conductance $g$ varies between 4.9 and 5.03 for the orthogonal model, and increases from 7 to 11 
for the S model). 
Linear fits shown by solid lines confirm that both $K_{11}$ and $K_{12}$ $\sim 1/L$  and
$\gamma$ equals to the symmetry parameter $\beta$ in the diffusive regime.
The spatial distribution of electrons is homogeneous and 
no additional parameter  must be introduced into the model. 
The transport is universal, the only model parameter
in  the DMPK  is the ratio $L_z/\ell$ of the system length to the mean free path.
Although the DMPK was derived only for the quasi-one dimensional systems, 
our data show that relations (\ref{dmpk-q1do},\ref{dmpk-q1du},\ref{dmpk-q1ds}) 
are valid also for the square samples.

\begin{figure}[t!]
\includegraphics[width=6.0cm]{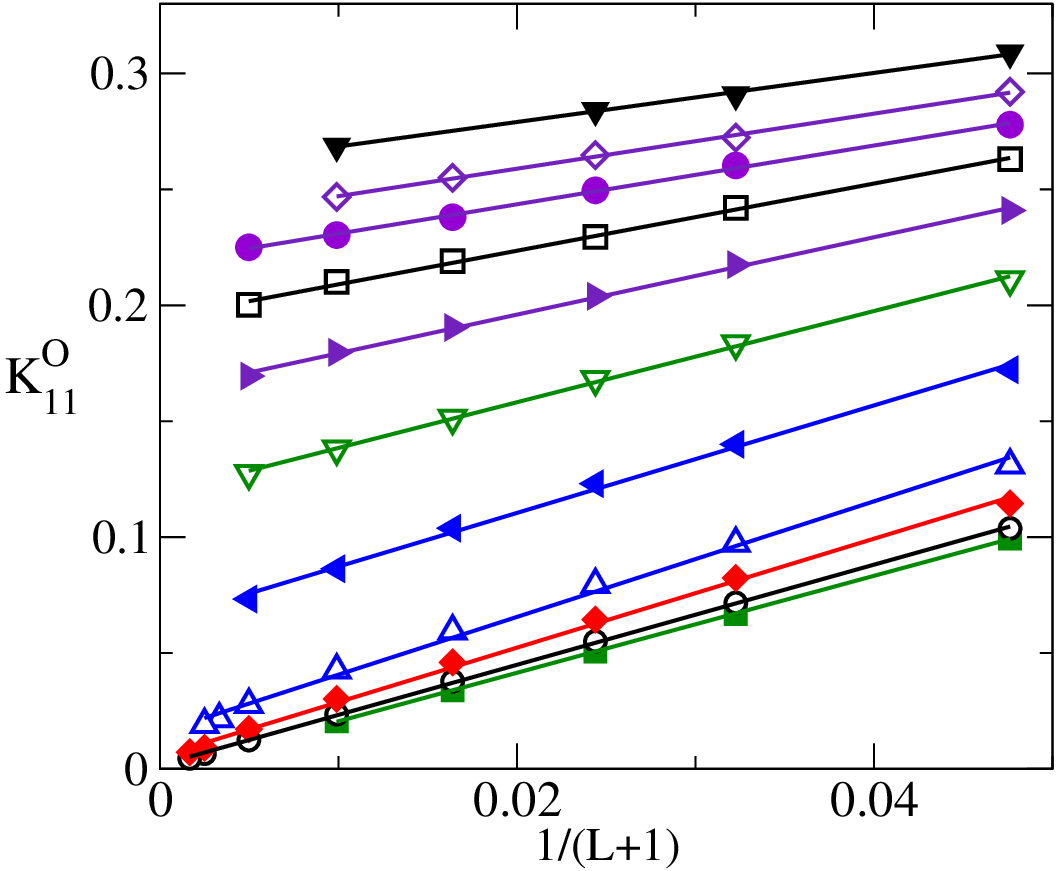}\\
\vspace*{1cm}
\includegraphics[width=6.0cm]{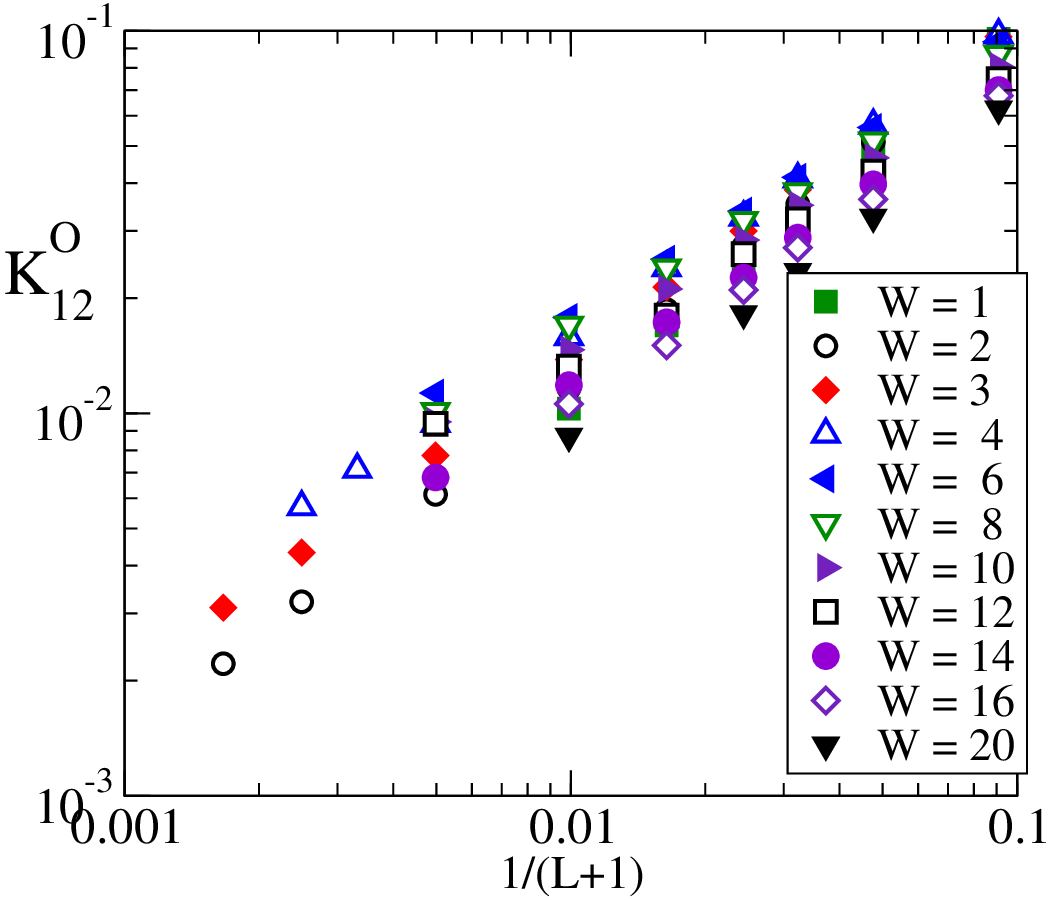}
\caption{(Color online) The size dependence of $K_{11}$ (top)   and $K_{12}$ (bottom)  for the 2D orthogonal model for various strength of the disorder (given in the legend of bottom panel).
$K_{11}$ converges 
to the non-zero limit when $L\to\infty$ for each disorder strength. This limiting value, however, is 
too small to be observable numerically for weak disorder within the considered size of the system. 
$K_{12}$ decreases to zero for any value of the disorder $W$.
}
\label{2D_K11_x}
\end{figure}

\begin{figure}
\includegraphics[width=6.0cm]{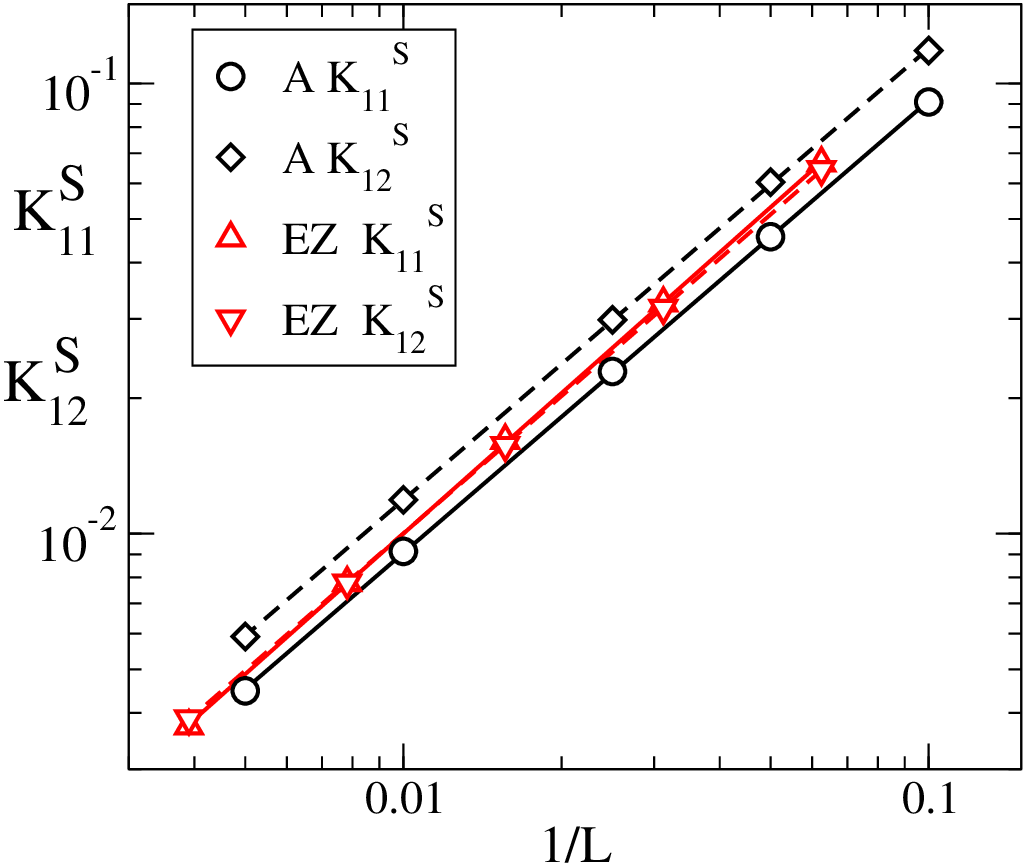}
\caption{(Color online) The size dependence 
$K_{11}$ and $K_{12}$   at the critical point for the 
symplectic (Ando and EZ model)   models.
Solid and dashed lines are power fits $K_{1a}\propto L^{-\alpha}$ ($a=1,2$)
with the exponent $\alpha\approx 1.003 - 1.005$.
}
\label{ando-cp}
\end{figure}

\begin{figure}
\vspace*{1cm}
\includegraphics[width=8.0cm]{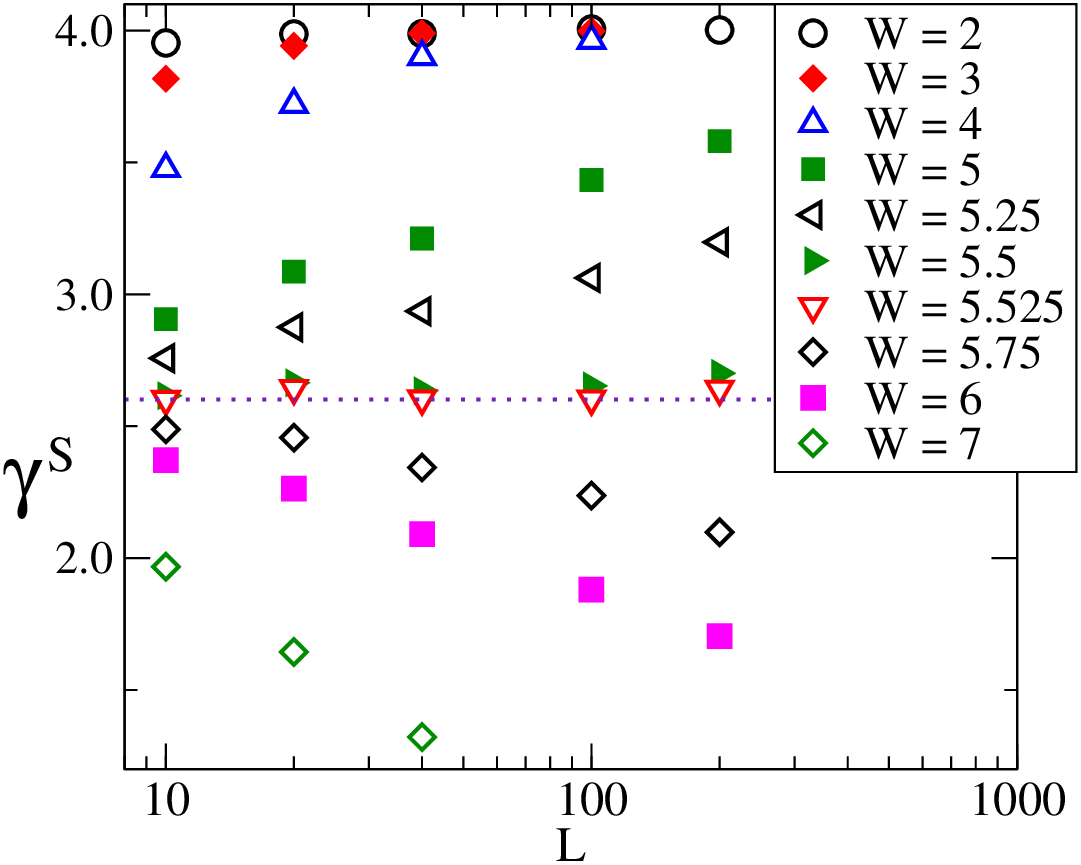}
\caption[]{(Color online)  The size dependence  of the parameter
 $\gamma^{\rm S}=2 K^{\rm S}_{12}/K^{\rm S}_{11}$ for the Ando model.
In the metallic regime, ($W<W_c=5.525$),   $\gamma^{S}$ slowly increases  when $L$ increases 
and converges to its isotropic value $\gs = 4$. 
In the localized regime ($W>W_c=5.525$),
$\gamma^{\rm S}$ decreases  to zero  when the size of the system increases.
At the critical point, $\gs$ does not depend on $L$.
The $L$-independent critical 
value of $\gamma^{\rm S}\approx 2.60$ (dotted line) is plotted by dot line. 
}
\label{gamma}
\end{figure}

\begin{figure}
\vspace*{1cm}
\includegraphics[width=6.0cm]{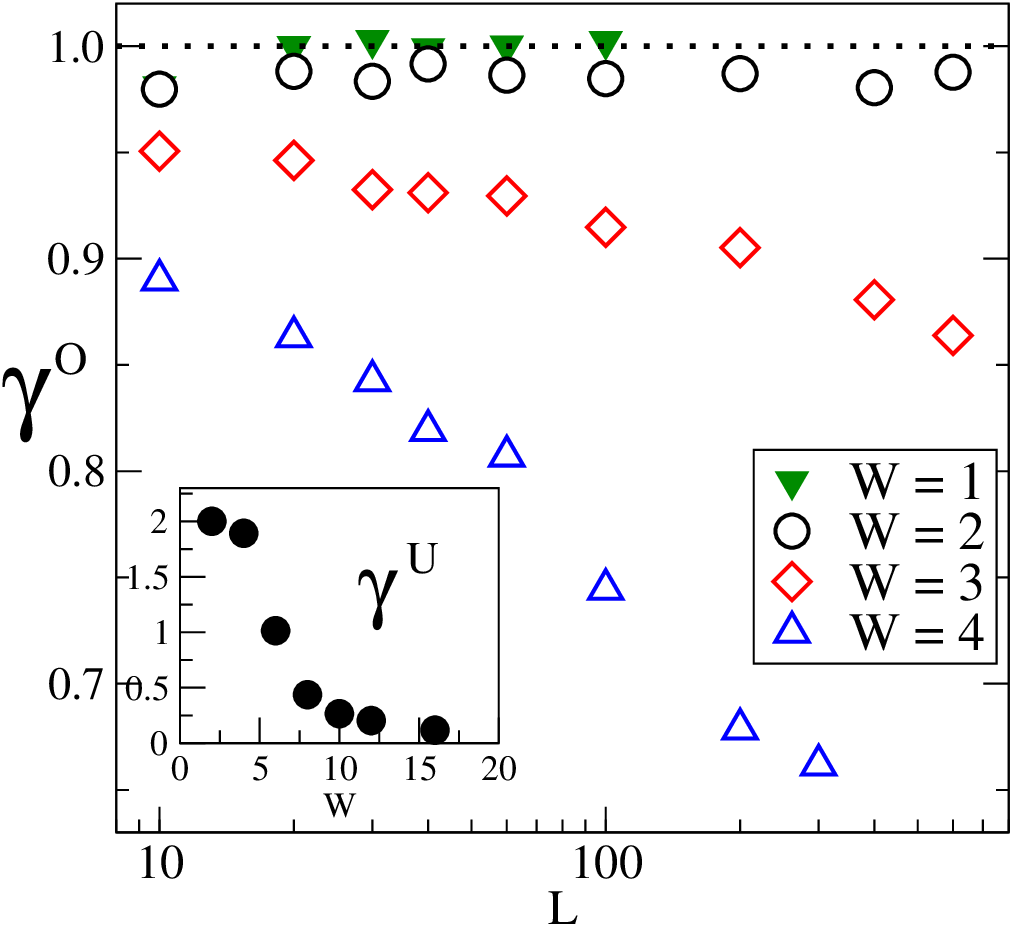}
\caption{(Color online) The size dependence  of the parameter $\gamma^{\rm O}=2 K^{\rm O}_{12}/K^{\rm O}_{11}$ for the 2D orthogonal  model.
We expect that $\gamma^{\rm O}$ decreases to zero for any value of the disorder $W$. However, the size of 
the system $L$ is not sufficient to see this decrease when the disorder is small. 
Instead, we observe the diffusive limit  $\gamma^{\rm O} \approx 1$. 
Only for sufficiently strong disorder, $\gamma^{\rm O}$
 decreases to zero. 
In contrast to the symplectic model (Fig. \ref{gamma}) there is no indication for the existence of the critical
point where $\gamma^{\rm O}$ converge to the $L$-independent limit.
Inset shows the disorder dependence of $\gamma^{\rm U}$ for the unitary ensembles ($L=128$, $\alpha=1/8$).
 }
\label{gamma-ortho}
\end{figure}

\subsection{Insulating   regime}

In the limit of strong disorder,
we expect that  $K_{aa}$  depend on the index $a$
and  $K_{aa} \sim O(1)$. Contrary, off-diagonal elements
$K_{ab}$, $a\ne b$, should decrease to zero,  $K_{ab} \sim 1/L$ ($a\ne b$)
so that $\gamma\sim 1/L$ decreases to zero when the size of the system increases.\cite{mk}

Figure \ref{2D_K11_x} shows the $L$ dependence of $K^{\rm O}_{11}$ and $K^{\rm O}_{12}$ for
orthogonal systems with various strength of the disorder.
Similarly to the 3D orthogonal model discussed in \cite{MMW}, both $\KO_{11}$ and 
$\KO_{12}$ are linear functions of $1/(L+1)$,
Since no metallic regime exists for the non-zero disorder, we expect that
$\KO_{11}$ converges to the   nonzero value in the limit
of $L\to\infty$ for all values of $W$,
\be\label{koko}
\KO_{11} = {\KO_{11}}^{\infty} + \ds{\frac{c}{L+1}}.
\ee
The limiting value
${\KO_{11}}^{\infty}$ can be easily calculated numerically for strong disorder. 
This is more difficult 
for  weak  disorder ($W< 4$), since   ${\KO_{11}}^{\infty}$ becomes
smaller  than the  inverse of the accessible sample size.

Similar data (not shown) were obtained also for the symplectic models.

\subsection{Critical regime (symplectic models)}\label{CR}

Critical regime exists only for the S systems.
In the critical regime, $W=W_c$  we found  that 
both $\KS_{11}$ and $\KS_{12}$ decreases at the critical point to zero 
\be
\KS_{11}(W=W_c),~~\KS_{12}(W=W_c)\propto \ds{\frac{1}{L}}
\ee
(Fig. \ref{ando-cp}), so that $\gs$ reaches a critical value,
$\gs_c=2\KS_{12}/\KS_{11}$ which does  not depend on the size of the system
\be
\gs_c = {\rm const}.
\ee
As shown in Fig. \ref{ando-cp}, the critical value
$\gamma_c^{\rm S}$  is not universal but depend on the  model.
We obtain $\gs = 2.601$ for the Ando model, and 1.795 for the Evangelou-Ziman model.

Figure \ref{gamma} shows that the length and disorder dependence
of parameter  $\gs$ can be, at least in principle,
used for the estimation of critical parameters in the same way as mean conductance
of the smallest Lyapunov exponent.
For very weak disorder, we find  that $\gs$ only weakly depends
on the size of the system and  increases to the metallic limit $\gamma= 4$ when $L$ increases to
infinity, indicating that  the system is in the metallic regime.
For stronger disorder,  $\gamma^{S}\propto 1/L$ decreases
 to zero when the size of the system increases, in agreement with the prediction of the theory.\cite{mk}
We found the critical regime between these to limits, 
where $\gamma^{\rm S}$ converges to the size-independent constant
$\gamma^{\rm^S}_c\approx 2.60$ (obtained already in Fig. \ref{ando-cp})  when $W=W_c$.

For comparison, we show in Fig. \ref{gamma-ortho}
 data for $\gamma^{\rm O}$  calculated for the 2D orthogonal model.
We found no critical regime.  Although $\gamma^{\rm O}\approx 1$ for weak disorder,
we expect that this is the finite size effect, and $\gamma^{\rm O}$ will decrease to zero for each disorder strength
when $L$ increases.\cite{AALR} 


\begin{figure}
\includegraphics[width=6.0cm]{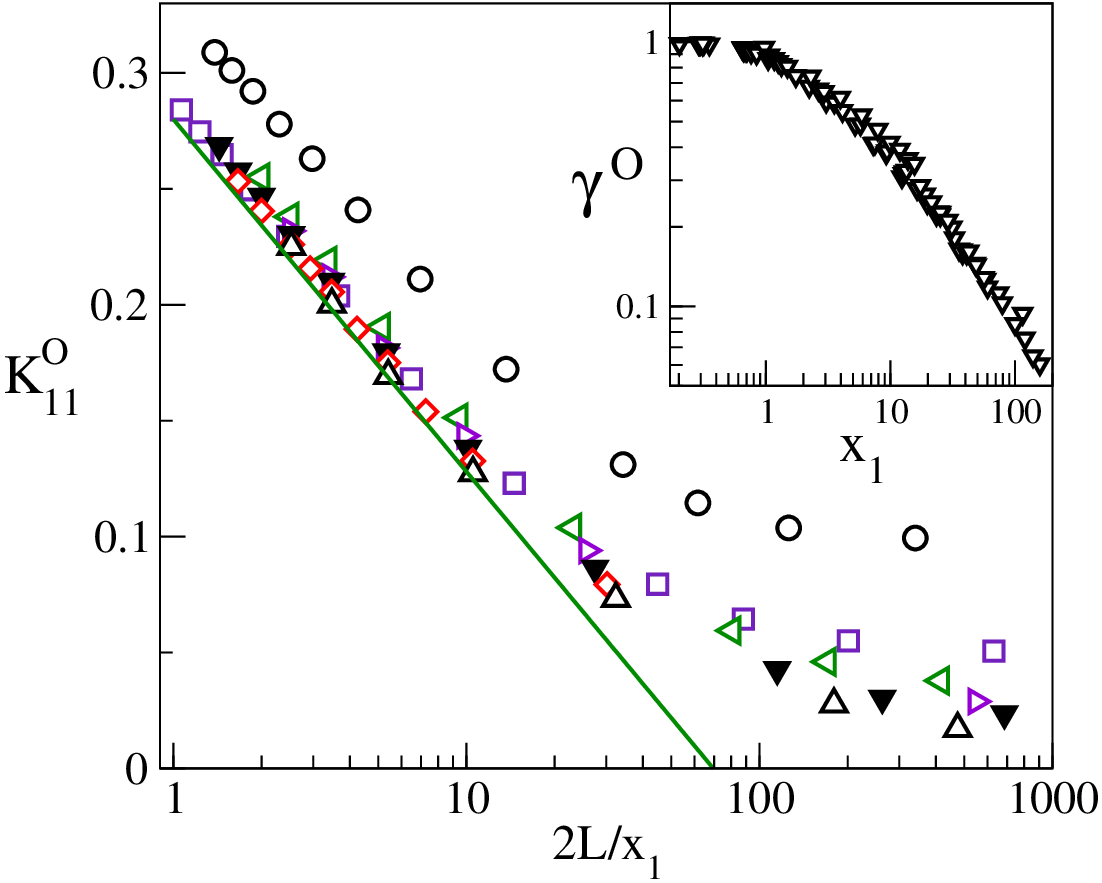}\\
\vspace*{1cm}
\includegraphics[width=6.0cm]{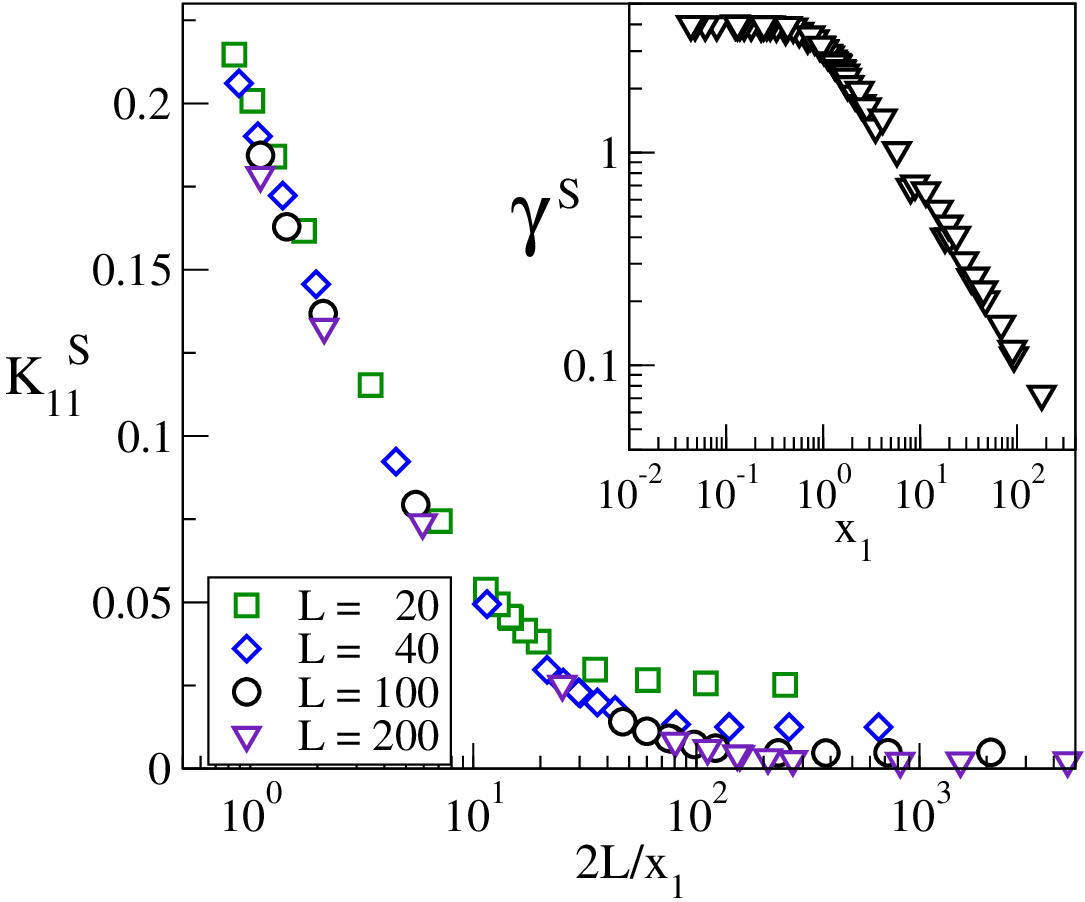}
\caption{(Color online) $K_{11}$ as a function of the localization length $\xi$ (estimated as $\xi=2L/x_1$) 
for the 2D orthogonal model (top)  and the 2D symplectic model (bottom).
The strength of the disorder varies between $W=1$ and $W=16$ in both Figures.
For strong disorder ($2L/x_1\to 0$) data converge to the universal curve 
(solid line is a function $0.288-0.066\ln\xi$ for the orthogonal model). 
Insets in  both panels show  that $\gamma$ is an unambiguous  function of $x_1$.
}
\label{xi_K11}
\end{figure}

\subsection{The universality}

With two  new parameters $K_{11}$ and $\gamma$, we must verify
if the transport properties of the system  are
  still maintained by only a single parameter.\cite{AALR} 
In the metallic regime, the answer is trivial since the entire matrix $K$ reduces to 
model-independent numbers given by Eqs. (\ref{dmpk-q1do},\ref{dmpk-q1du},\ref{dmpk-q1ds}).
The universality of the critical regime was shown in the previous section.
Here, we concentrate on the localized regime, where 
we expect that $K_{11}$ becomes an unambiguous function of the localization length. \cite{MMW}
The last can be estimated
from the smallest parameter $x_1$, 
\be
\xi=\ds{\frac{2L}{x_1}}.
\ee
In  Fig. \ref{xi_K11}  we plot
$K_{11}$ as a function of $\xi$ for the orthogonal and symplectic Ando 
models. Data confirm that  the parameter $K_{11}$ becomes an
linear  function of $\ln \xi$  with increasing system size  and converge to the
system-size independent limit when $L\to\infty$.

Two inset  of Fig. \ref{xi_K11}, show that the parameter
$\gamma$ is an unambiguous function of $x_1$ in all three regimes. 
In the localized regime, when $x_1\sim L$,  data confirm that $\gamma\sim 1/L$, 
consistent with prediction of the Muttalib's theory.

\begin{figure}
\includegraphics[width=8.0cm]{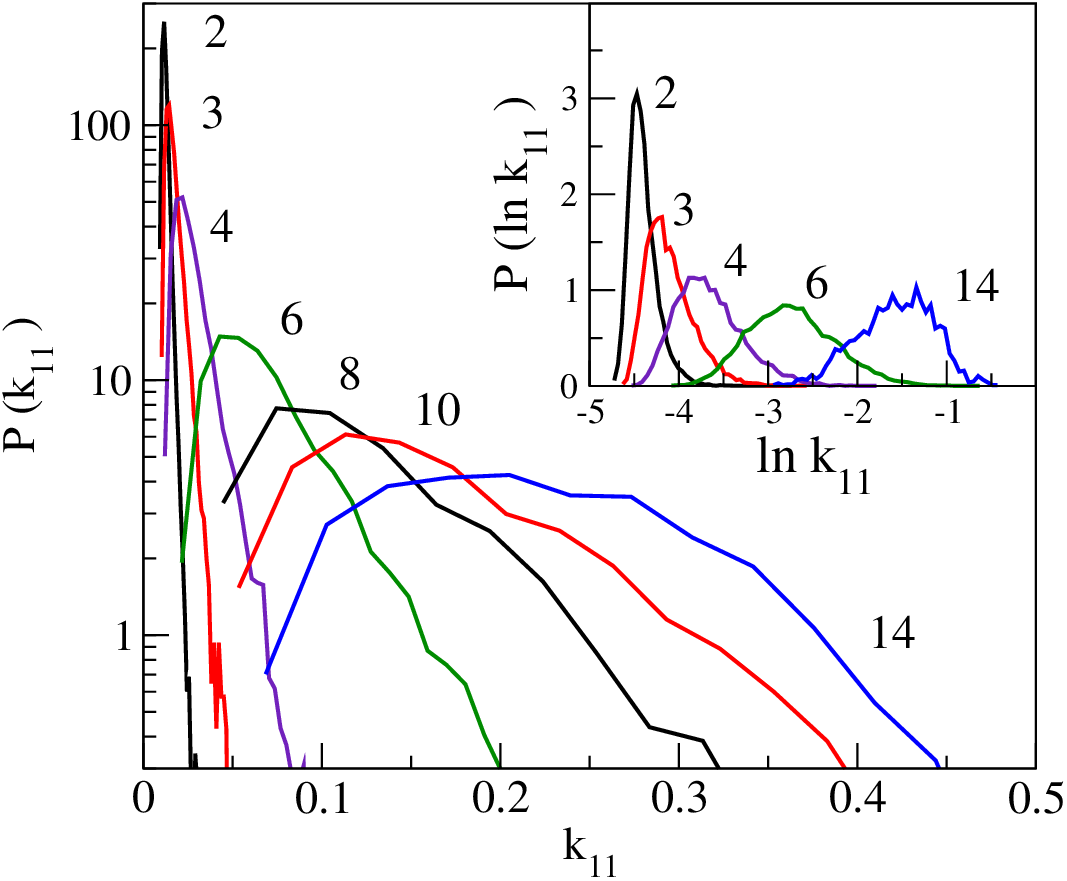}
\caption[]{(Color online) The  probability distribution of $k_{11}$ for the 2D orthogonal model
with  various  disorder strength (given in the figure). 
The size of the system $L=200$. Inset shows the distributions $P(\ln K_{11})$.
Data confirm that $k_{11}$ is a good statistical variable with a  well defined mean value and variance.
}
\label{K11_2D}
\end{figure}

\begin{figure}
\vspace*{1cm}
\includegraphics[width=6.0cm]{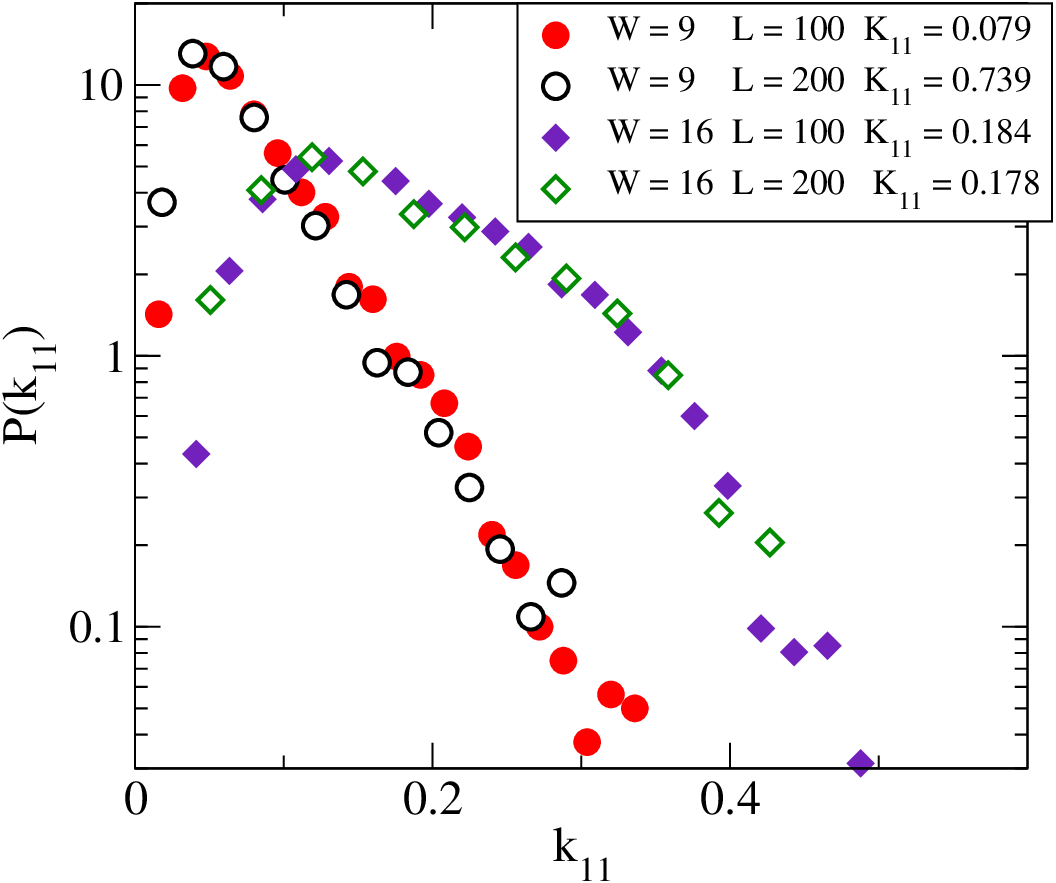}
\caption{(Color online) The  probability distribution of $k_{11}$ in the strongly localized 
regime for the Ando model. 
The mean value  $K_{11}$ is given in the legend.
}
\label{K11_2D_w14_L}
\end{figure}

\begin{figure}[t!]
\includegraphics[width=6.0cm]{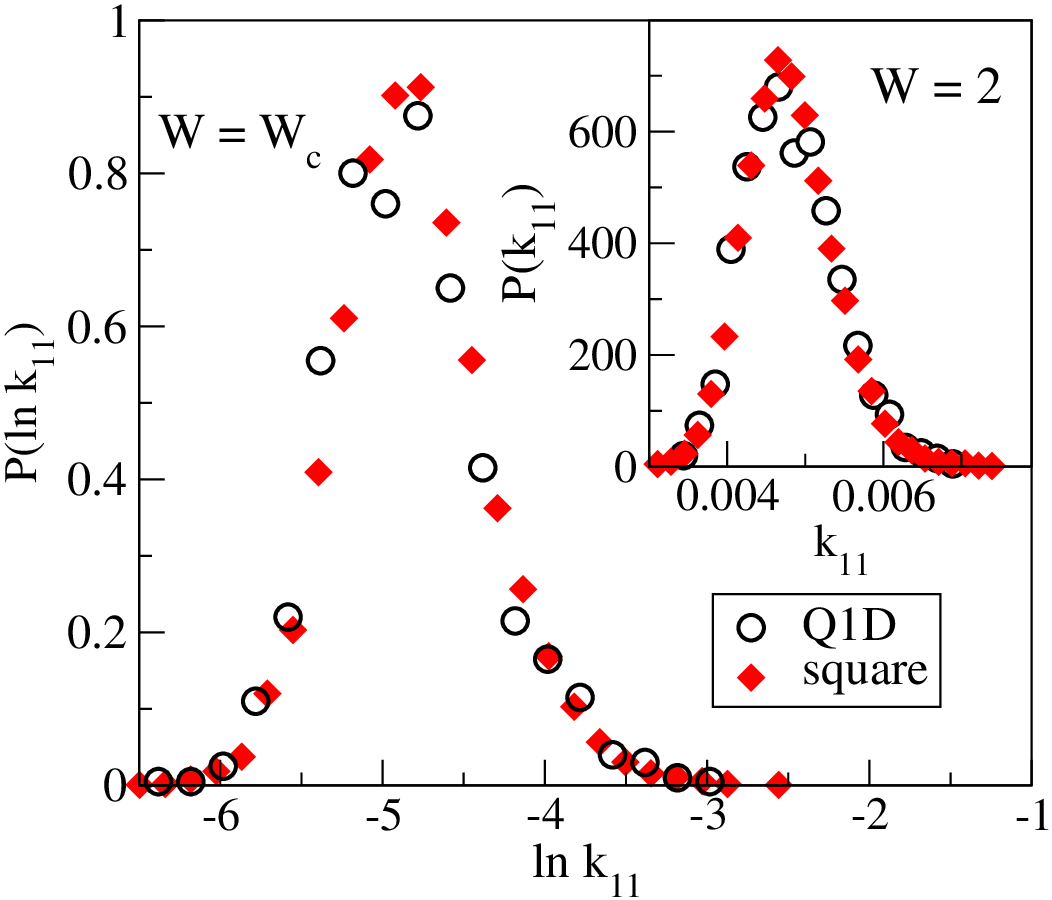}
\caption{(Color online) The  probability distribution of $k_{11}$ for the 2D Ando model 
at the critical point $W=W_c = 5.525$
and in the metallic regime ($W=2$, inset).
Both  the square samples $100\times 100$  and Q1D samples $100\times 5000$ are considered.
}
\label{ando-pk11}
\end{figure}

\subsection{Statistical properties of $k_{11}$}

In the previous analysis we dealt only with mean values of $K_{11}$ and $K_{12}$. Since both
$k_{11}$ and $k_{12}$ are statistical variables,  we must  also to study their statistical properties.
Figure \ref{K11_2D} shows the probability distribution of  parameter $k_{11}$  and $\ln k_{11}$ for the
2D orthogonal model. For each disorder, the mean value can be identified with the most probable 
value. 
In the localized limit, both  $K_{11}$ and var~$k_{11}$ are of order of unity, 
and  the distribution $P(k_{11})$ becomes size independent 
(Fig.  \ref{K11_2D_w14_L}). 

In Fig. \ref{ando-pk11} we plot the probability distribution of $k_{11}$ for the symplectic Ando model
in the critical and metallic regime.  We demonstrate  that 
the distributions for the square sample $L\times L$ with quasi-one dimensional systems are almost identical.


\begin{figure}[t]
\includegraphics[width=6.0cm]{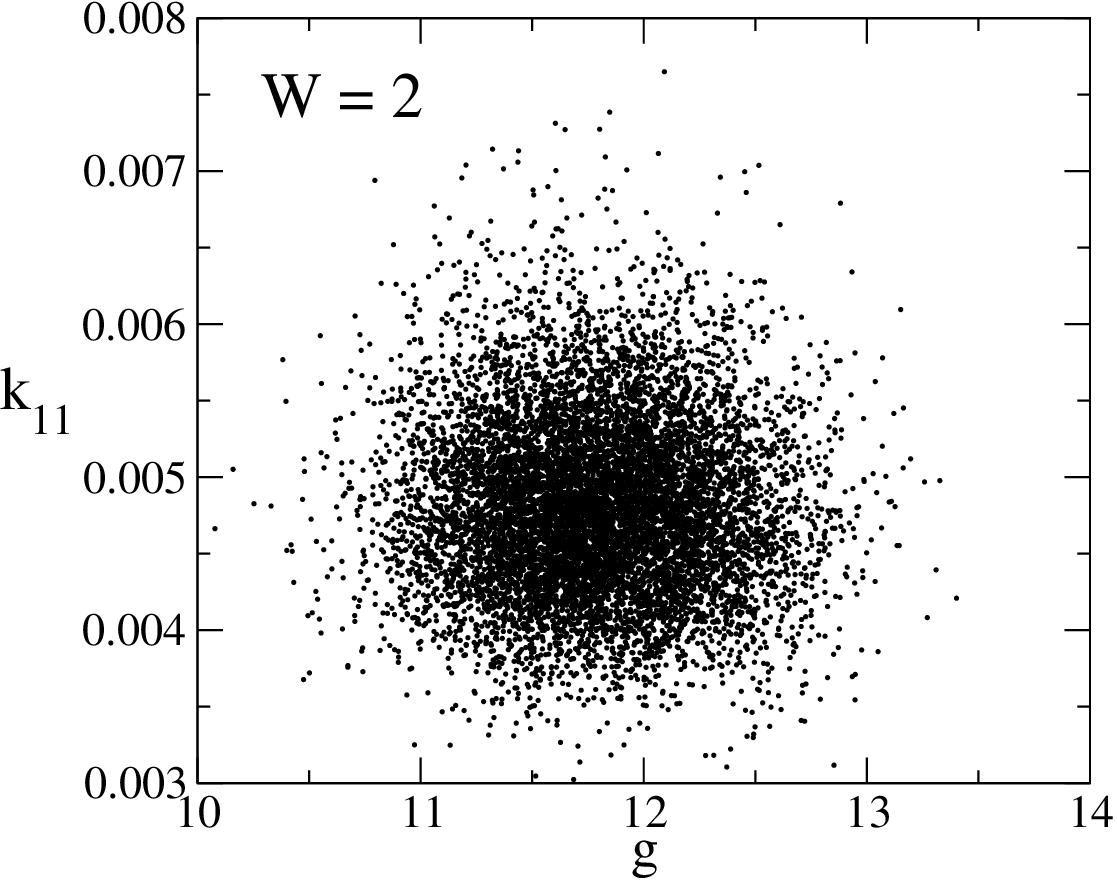}\\
\vspace*{2cm}
\includegraphics[width=6.0cm]{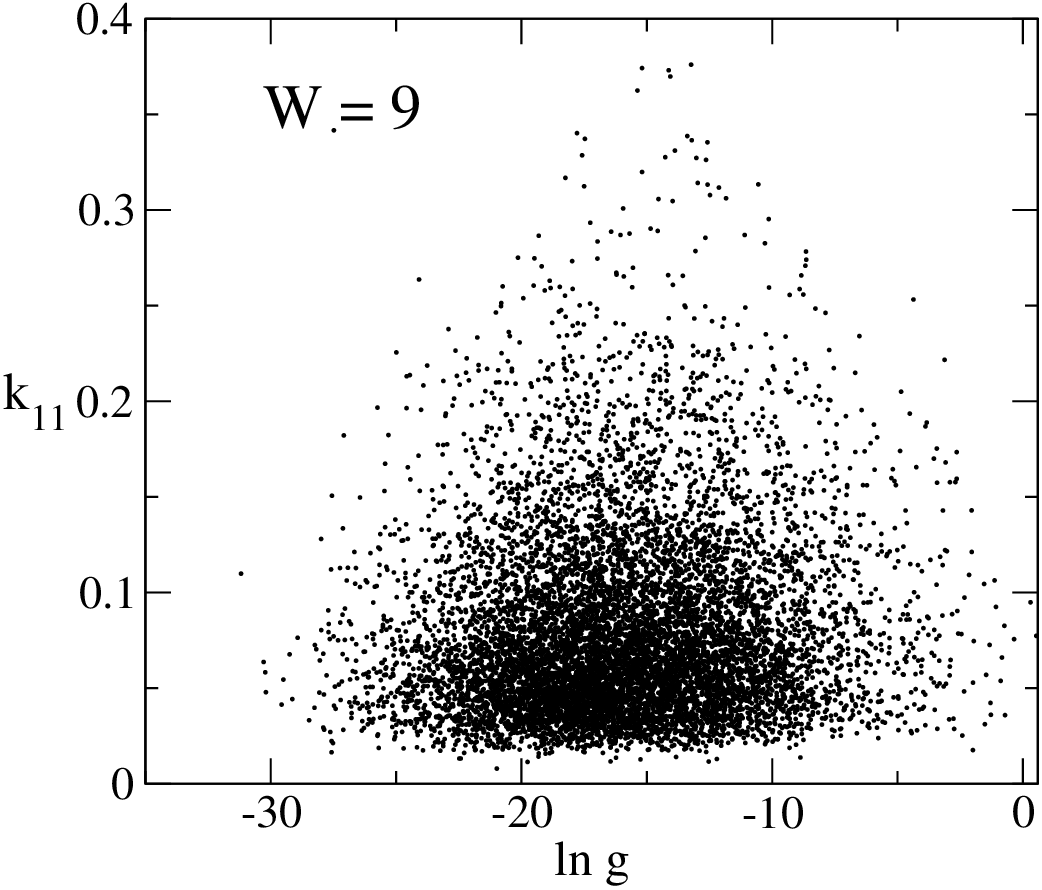}
\caption[]{2D Ando model: $k_{11}$ as a function of the conductance
statistical ensemble of  $N_{\rm stat}=10^4$ samples with disorder
$W=2$ (top) and $W=9$ (bottom). The size of the system $L=100$. 
}
\label{g_K11}
\end{figure}

\subsection{Correlation $g$ vs $k_{11}$}

We have shown that $K_{11}\to 0$ in the metallic regime but $K_{11}\sim O(1)$ in the insulator.
Small values of $K_{11}$ indicate that the mean conductance of the system is large. 
Contrary, large values of $K_{11}$ correspond to systems with small mean conductance.
This is in agreement with our expectation: small conductance means that the electron has problems
to go through the sample. When it finally reaches the opposite side, its spatial distribution
is not homogeneous any more.

However, the correspondence large $k_{11}$ - small $g$ holds only for mean values of these parameters.
As shown in Fig. \ref{g_K11}, the values of $g$ and $k_{11}$ 
for a given sample are not correlated within a given statistical ensemble:
small values $g\ll \langle g\rangle$ can be accompanied with any value of $k_{11}$ - 
either small $k_{11}\ll K_{11}$ or large $k_{11}\gg K_{11}$. 
The absence  of the  correlation observed in 
both  the metallic and in strongly localized regime, confirms that the statistical fluctuations 
of $k_{11}$ do not affect the mean value of the conductance.

\section{Conclusion}

The 
electron transport through disordered system is determined by spatial 
distribution of the electron inside the disordered sample,
which can be measured by parameters $K_{11}$ and $\gamma$. Our aim in this paper was to investigate
how these two parameters depend on the size of the system, strength of the disorder and physical symmetry
of the model. We concentrated on 2D disordered systems. 
In order to better
understand the role of the disorder, we compare numerical data for the orthogonal and symplectic
physical symmetry. 
For completeness, we add also a few data for the unitary ensemble.

In the diffusive regime, the size dependence of both parameters follow the analytical relations
given by the theory of DMPK equation. In particular, $\gamma$ equals to the symmetry parameter $\beta$.
In the localized regime, $K_{11}$ converges to the size independent limit and
$\gamma\sim 1/L$.  

For  the symplectic models, which  exhibit the metal-insulator transition,
we  analyze the size dependence of both parameters 
and we found that $\gamma$ possesses a critical value $\gamma_c$ when disorder $W=W_c$. 
Also, we found no significant difference between the values of $K$ 
for the two dimensional and quasi-one dimensional systems.
No critical value was found for the orthogonal model.

We also found that $K_{11}$ is an unambiguous function of the localization length $\xi$ and
$\gamma$ is uniquely given by the parameter $x_1$. Therefore, the use of these parameters does not 
contradict the single parameter scaling theory.

Since the elements of matrices  $k$ are  given by elements of statistical matrices $v$,  
they are also statistical variables.
Fortunately, analysis of their probability distributions confirm that
their mean values are good representatives of the statistical ensembles.
We found no statistical correlations between the conductance and $k_{11}$.
Therefore, we conclude that mean values, $K_{11}$ and $K_{12}$, and, 
consequently, $\gamma=2K_{12}/K_{11}$, are  physical parameters for the description of disordered systems.

\bigskip

Acknowledgments: This work was supported by  project VEGA 0633/09.


\begin{thebibliography}{99}
\bibitem{comment} In the absence of the disorder, the $z$-component of the wave vector
$$
\cos k_za = \ds{\frac{1}{2V_\parallel}}[E-2V_\perp\cos k_xa]
$$
is real for all possible values of $k_x$ provided that the energy $E$ is close to the
band center $E=0$.
\bibitem{xx} For symplectic models, each element of $M$ and $M_0$ is the $2\times 2$ matrix.
\end{thebibliography}
\end{document}